\documentclass[prd,preprint,tightenlines,floatfix,showpacs,preprintnumbers,nofootinbib,eqsecnum]{revtex4}

 \usepackage[dvips,final]{graphicx}
  \usepackage{amssymb}
   \usepackage{amsmath}
    \usepackage{amsfonts}
     \usepackage{epsfig}
      \usepackage{bm}

\begin{document}

\thispagestyle{empty} \preprint{\hbox{}} \vspace*{-10mm}

\title{$pp \to pp\eta'$ reaction at high energies
\footnote{Dedicated to Andrzej Bia{\l}as on the occasion of his 70th birthday}}

\author{A.~Szczurek}

\email{antoni.szczurek@ifj.edu.pl}

\affiliation{Institute of Nuclear Physics PAN, PL-31-342 Cracow,
Poland} \affiliation{University of Rzesz\'ow, PL-35-959 Rzesz\'ow,
Poland}

\author{R.~S.~Pasechnik}

\email{rpasech@theor.jinr.ru}

\author{O.~V.~Teryaev}

\email{teryaev@theor.jinr.ru}

\affiliation{ Bogoliubov Laboratory of Theoretical Physics, JINR,
Dubna 141980, Russia} \affiliation{ Faculty of Physics, Moscow
State University, Moscow 119992, Russia}

\date{\today}

\begin{abstract}
We discuss double-diffractive (double-elastic) production of the
$\eta'$-meson in the $pp \to p\eta'p$ reaction within the
formalism of unintegrated gluon distribution functions (UGDF). We
estimate also the contribution of $\gamma^* \gamma^* \to \eta'$ fusion.
The distributions in the Feynman $x_F$ (or rapidity), transferred 
four-momenta squared between initial and final protons ($t_1$, $t_2$)
and azimuthal angle difference between outgoing protons ($\Phi$)
are calculated.
The deviations from the $\sin^2(\Phi)$ dependence predicted by
one-step vector-vector-pseudoscalar coupling are quantified and discussed.
The results are compared with the results of the WA102
collaboration at CERN. 
Most of the models of UGDF from the literature
give too small cross section as compared to the WA102 data
and predict angular distribution in relative azimuthal angle strongly
asymmetric with respect to $\pi/2$ in disagreement with the WA102 data.
This points to a different mechanism at the WA102 energy.
Predictions for RHIC, Tevatron and LHC are given.
We find that the normalization, $t_{1,2}$
dependences as well as deviations from $\sin^2(\Phi)$ of
double-diffractive double-elastic cross section are extremely sensitive
to the choice of UGDF. Possible implications for UGDFs are discussed.
\end{abstract}



\maketitle

\section{Introduction}

The search for Higgs boson is the primary task for the LHC collider
being now constructed at CERN. Although the predicted cross section
is not small it may not be easy to discover Higgs in inclusive
reaction due to large background in each of the final channel considered.
An alternative way \cite{SNS90,BL91,CH96} is to search for Higgs
in exclusive or semi-exclusive reactions with large rapidity gaps.
Although the cross section is not large, the ratio of the signal to
more conventional background seems promising.
Kaidalov, Khoze, Martin and Ryskin proposed to calculate diffractive
double elastic \footnote{Both protons survive the collision.} production
of Higgs boson in terms of unintegrated gluon distributions
\cite{KMR97,KMR02,KKMR03,KKMR04}. It is not clear at present how
reliable such calculations are. It would be very useful to use the
formalism to a reaction which is easy to measure. Here we shall
try to apply it to the production of $\eta'$ meson which 
satisfies this criterium.

Recently the exclusive production of $\eta'$ meson in
proton-proton collisions was intensively studied slightly above
its production threshold at the COSY ring at KFA J\"ulich
\cite{COSY11} and at Saclay \cite{DISTO}. Here the dominant
production mechanism is exchange of several mesons (so-called
meson exchange currents) and reaction via $S_{11}$ resonance
\cite{COSY_theory}.

In the present note we study the same exclusive channel but at
much larger energies ($W >$ 10 GeV). Here diffractive
mechanism is expected to be the dominant process.
In Ref.\cite{KMV99} the Regge-inspired pomeron-pomeron fusion was
considered as the dominant mechanism of the $\eta'$ production.

There is a long standing debate about the nature of the pomeron.
The approximate
$\sin^2(\Phi)$ ($\Phi$ is the azimuthal angle between outgoing
protons) dependence observed experimentally \cite{WA102} was
interpreted in Ref.\cite{CS99} as due to (vector pomeron)-(vector
pomeron)-(pseudoscalar meson) coupling. To our knowledge no
QCD-inspired calculation for diffractive production of
pseudoscalar mesons exists in the literature.


Of course, one should worry about the origin of the hard scale 
which may justify the applicability of QCD perturbation theory.
As soon as the mass of  $\eta'$ is not sufficient for that, 
it can be reasonably large $\eta'$  transverse momentum (or $t_{1,2}$)
which would serve as a hard scale. Bearing this in mind, we will examine 
the QCD result in the whole kinematical region, which should be
understood as a sort of continuation of perturbative result to
the region where its applicability cannot be rigorously proven.


\begin{figure}[!h]    
 \centerline{\includegraphics[width=0.5\textwidth]{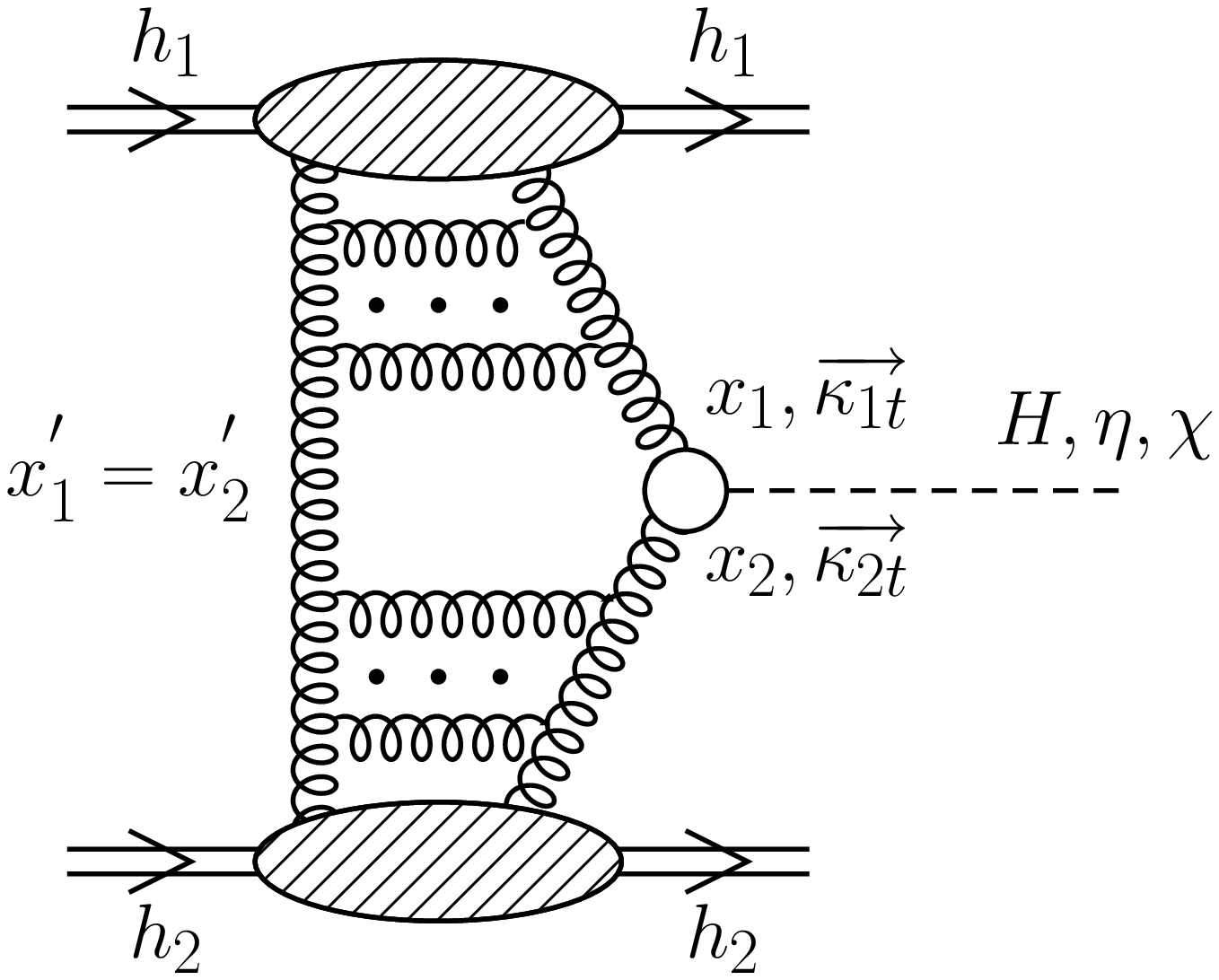}}
   \caption{\label{fig:diffraction_updf}
   \small  The sketch of the bare QCD mechanism. The kinematical
variables are shown in addition.}
\end{figure}

\begin{figure}[!h]    
 \centerline{\includegraphics[width=0.3\textwidth]{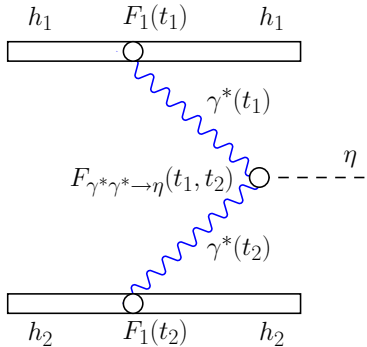}}
   \caption{\label{fig:gamgam_eta}
   \small  The sketch of the photon-photon fusion mechanism.
Form factors appearing in different vertices are shown explicitly.}
\end{figure}

In Fig.\ref{fig:diffraction_updf} we show the QCD mechanism of diffractive
double-elastic production of $\eta'$ meson, analogous to the mechanism
of Higgs boson production. We shall show here
that approximate ($\sim \sin^2(\Phi)$) dependence is violated
in the QCD-inspired model with gluon exchanges within the formalism
of unintegrated gluon distribution functions (UGDF) and a
distortion from this dependence can help to select the correct
model of UGDF. For completeness, in this paper we shall include
photon-photon fusion mechanism shown in Fig.\ref{fig:gamgam_eta} which
was sometimes advocated as dominant mechanism at high energies.


\section{Formalism}

\subsection{Diffractive QCD mechanism}


\begin{figure}[h]     
 \centerline{\includegraphics[width=0.6\textwidth]{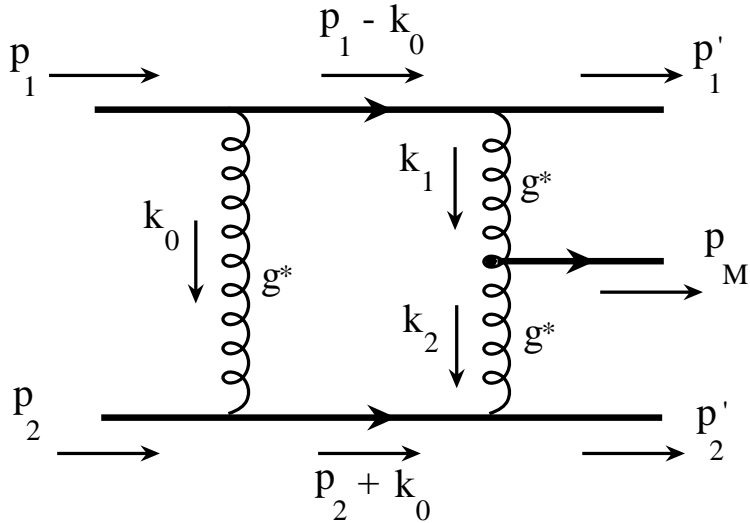}}
   \caption{\label{kinematics_qcd}
   \small Kinematics of exclusive double-diffractive $\eta'$-meson production.}
\end{figure}



The kinematics of the process on the quark level is
shown in Fig.~\ref{kinematics_qcd}.
The decomposition of gluon momenta into longitudinal and
transverse parts gives
\begin{eqnarray}
k_0=-x'_1p_1+k_{0,t}=x'_2p_2+k_{0,t},\quad
k_1=x_1(k_0-p_1)+k_{1,t},\quad k_2=x_2(p_2+k_0)+k_{2,t}\, .
\label{dec}
\end{eqnarray}

We take into account below that $x'_1=x'_2=x_0$. Making use of
conservation laws we get
\begin{eqnarray}
k_1+p'_1=p_1-k_0,\qquad k_2+p_2+k_0=p'_2\, . \label{CL}
\end{eqnarray}
Taking the transverse parts from these relations gives
\begin{eqnarray}
k_{1,t}=-(p'_{1,t}+k_{0,t}),\qquad k_{2,t}=p'_{2,t}-k_{0,t}\, .
\label{perp}
\end{eqnarray}

Following the formalism for the diffractive double-elastic
production of the Higgs boson developed by Kaidalov, Khoze, Martin and Ryskin
\cite{KMR97,KMR02,KKMR03,KKMR04} (KKMR) we write the bare QCD amplitude
for the process sketched in Fig.1 as \footnote{For a sketchy derivation of this
formula starting from the parton level one may look to
\cite{Forshaw05}.}
\begin{eqnarray}
{\cal M}_{pp \to p \eta' p}^{g^*g^*\to\eta'} =  i \, \pi^2 \int
d^2 k_{0,t} V(k_1, k_2, P_M) \frac{
f^{off}_{g,1}(x_1,x_1',k_{0,t}^2,k_{1,t}^2,t_1)
       f^{off}_{g,2}(x_2,x_2',k_{0,t}^2,k_{2,t}^2,t_2) }
{ k_{0,t}^2\, k_{1,t}^2\, k_{2,t}^2 } \, . \label{main_formula}
\end{eqnarray}
The normalization of this amplitude differs from the KKMR one
\cite{KMR02,KMRS04} by the factor $i$. The bare amplitude above is
subjected to absorption corrections which depend on collision
energy. We shall discuss this issue shortly when presenting our
results.

The vertex function $V(k_1,k_2,P_M)$ in the expression
(\ref{main_formula}) describes the coupling of two virtual gluons
to the pseudoscalar meson. We take the gluon-gluon-pseudoscalar
meson coupling in the form
\begin{eqnarray}
V_{\alpha\beta}(k_1,k_2,P_M)=V_N\,F_{g^*g^*\to\,\eta'}(k_1^2,k_2^2)\,\varepsilon_{\mu\nu\alpha\beta}\,k_1^{\mu}\,k_2^{\nu}\,.
\label{coupl}
\end{eqnarray}
Normalization is such that $F_{g^*g^*\rightarrow\,\eta'}(0,0)=1.$
The normalization constant $V_N$ can be obtained in terms of the
partial decay width $\Gamma(\eta'\rightarrow gg)$ as
\begin{eqnarray}
V_N^2=K\frac{64\pi\Gamma(\eta'\rightarrow
gg)}{(N_c^2-1)m_{\eta'}^3},\qquad
\textrm{NLO}\quad\rightarrow\quad K=1.5\,. \label{V_N_from_Gamma}
\end{eqnarray}
The same normalization was obtained for the QCD double-diffractive
production of $\chi$ mesons in \cite{KMRS04}.

The gauge invariance requires
$k_1^{\alpha}V_{\alpha\beta}=k_2^{\beta}V_{\alpha\beta}=0$. On the
parton level using Feynman rules we have to ``hook"
the gluon-gluon-pseudoscalar meson coupling (\ref{coupl}) to the quark
line by contracting it with incoming quark momenta. Replacing the
quark lines by the proton lines one can show that the vertex factor
$V(k_1,k_2,P_M)$ in the amplitude (\ref{main_formula}) has the
following form
\begin{eqnarray}
V=(k_0-p_1)^{\alpha}(p_2+k_0)^{\beta}V_{\alpha\beta}=
\frac{k_{1,t}^{\alpha}}{x_1}\frac{k_{2,t}^{\beta}}{x_2}
V_{\alpha\beta}    \; .
\label{vert}
\end{eqnarray}
Using relations (\ref{dec}),~(\ref{perp}) and (\ref{coupl}) this
expression can be transformed to
\begin{eqnarray*}
V=V_N\,F_{g^*g^*\to\,\eta'}(k_1^2,k_2^2)\,\varepsilon_{\mu\nu\alpha\beta}\,(p_1+p_2)^{\mu}\,(p_2+k_0)^{\nu}
(p'_{1,t}+k_{0,t})^{\alpha}(p'_{2,t}-k_{0,t})^{\beta}\, .
\end{eqnarray*}
In the c.m.s. system ${\bf p_1}+{\bf p_2}=0$ and
$(p_1+p_2)_0=\sqrt{s}.$ Since ${\bf k}_{0,t}\perp[{\bf
p'}_{1,t}\times{\bf p'}_{2,t}]$ we have
\begin{eqnarray}
V=-V_N\sqrt{s}\,F_{g^*g^*\to\,\eta'}(k_1^2,k_2^2)\,\varepsilon_{ikl}\,(p'_{1,t}+k_{0,t})_{i}\,
(p'_{2,t}-k_{0,t})_{k}\,p_{1,l}\, .
\label{V_midd}
\end{eqnarray}
Introducing a unit vector in the beam direction of ingoing protons
in c.m.s. $n_l=p_{1,l}/|{\bf p}_{1}|$, we get finally
\begin{eqnarray}
V=-V_N\frac{s}{2}\, F_{g^*g^*\to\,\eta'}(k_1^2,k_2^2)\,[({\bf
p}'_{1,t}+{\bf k}_{0,t})\times ({\bf p}'_{2,t}-{\bf
k}_{0,t})]\;{\bf n}\, . \label{V_fin}
\end{eqnarray}
Normalization of this vertex function differs from the KKMR one
\cite{KMR02,KMRS04} by the factor $s\,F_{g^*g^*\to\,\eta'}/2.$
Form factor $F_{g^*g^*\to\,\eta'}$ can be relevant at some
kinematical regions and it should be taken into account. Factor
$s/2$ makes the normalization of the full bare QCD amplitude
(\ref{main_formula}) consistent with canonical normalization of
the cross section (see Eq.(\ref{cross sect})).

Expression (\ref{V_midd}) can be also written as
\begin{eqnarray}
V(k_1,k_2,P_M) = V_N \frac{s}{2}\,F_{g^* g^* \to\,
\eta'}(k_1^2,k_2^2) \cdot |{\bf k}_{1,t}| |{\bf k}_{2,t}|  \cdot
\sin(\phi) \, , \label{gg_PS_vertex}
\end{eqnarray}
where $\phi$ is the azimuthal angle between ${\bf k}_{1,t}$ and
${\bf k}_{2,t}$. In our case, in contrast to vector-vector fusion
to pseudoscalars, $\phi \ne \Phi$. This may cause a deviation from
$\sin^2(\Phi)$ distribution. To better illustrate the deviation we
write the gluon-gluon-pseudoscalar meson coupling in the following
equivalent way
\begin{eqnarray}
V(k_1,k_2,P_M) =-V_N
\frac{s}{2}\,F_{g^*g^*\rightarrow\,\eta'}(k_{1,t}^2,
k_{2,t}^2)\, \nonumber \\
\times \biggl[|{\bf p'}_{1,t}||{\bf p'}_{2,t}|\sin(\Phi)-|{\bf
k}_{0,t}||{\bf p'}_{1,t}|\sin(\psi+\Phi)+|{\bf k}_{0,t}||{\bf
p'}_{2,t}|\sin(\psi) \biggr]\,. \label{vert_mod}
\end{eqnarray}
Here $\Phi$ is explicitly the azimuthal angle between the
transverse momenta of outgoing protons ${\bf p'}_{1,t}$ and ${\bf
p'}_{2,t},$ $\psi$ is the azimuthal angle between ${\bf k}_{0,t}$
and ${\bf p'}_{2,t}$ ($0<\psi<2\pi$).

It is convenient to introduce the dimensionless parameters
\begin{eqnarray*}
\mu=\frac{|{\bf k}_{0,t}|}{m_{\eta'}},\quad \xi=\frac{|{\bf
p'}_{1,t}|}{m_{\eta'}},\quad \eta=\frac{|{\bf
p'}_{2,t}|}{m_{\eta'}}\, .
\end{eqnarray*}
We take $k_{0,t}^2=-|{\bf k}_{0,t}|^2,$ $k_{1,t}^2=-|{\bf
k}_{1,t}|^2,$ $k_{2,t}^2=-|{\bf k}_{2,t}|^2$ and write
differential $d\,^2 k_{0,t}$ as: $d\,^2 k_{0,t}=-|{\bf
k}_{0,t}|d|{\bf k}_{0,t}|d\psi$. Then we obtain finally
\begin{eqnarray}
{\cal M}_{pp \to p \eta'
p}^{g^*g^*\to\eta'}(\xi,\eta,\Phi,m_{\eta'})=-V_N\,i\,\pi^2\,\frac{s}{2m_{\eta'}^2}
\int \frac{d\mu}{\mu}\times \nonumber
\\
\int_0^{2\pi}
d\psi\,\frac{[\xi\eta\sin(\Phi)-\mu\xi\sin(\psi+\Phi)+\mu\eta\sin(\psi)]}{[\mu^2+\xi^2+2\mu\xi\cos(\psi+\Phi)]
[\mu^2+\eta^2-2\mu\eta\cos(\psi)]}\times \label{matrelem}
\\
\times f^{off}_{g,1}(x_1,x_1',k_{0,t}^2,k_{1,t}^2,t_1)
f^{off}_{g,2}(x_2,x_2',k_{0,t}^2,k_{2,t}^2,t_2)F_{g^*g^*\rightarrow\,\eta'}(k_{1,t}^2,
k_{2,t}^2) \; . \nonumber
\end{eqnarray}
We have to take into account that the dimensionless arguments in
$f^{off}_{g,1},$ $f^{off}_{g,2}$ and
$F_{g^*g^*\rightarrow\,\eta'}$
\begin{eqnarray*}
& &{}
\frac{k_{0,t}^2}{m_{\eta'}^2}=-\mu^2,\;\frac{k_{1,t}^2}{m_{\eta'}^2}=-\mu^2-\xi^2-2\mu\xi\cos(\psi+\Phi),
\\
& &{}
\frac{k_{2,t}^2}{m_{\eta'}^2}=-\mu^2-\eta^2+2\mu\eta\cos(\psi)
\end{eqnarray*}
are the functions of integration variables $\mu$ and $\psi.$ So
now we clearly see there is no simple angular behavior like ${\cal
M}\sim\sin(\Phi)$. The angular behavior of matrix element is more
complicated. Only in the limit $k_{0,t}\rightarrow 0$ the
$\sin(\Phi)$-behavior with some modulated amplitude is restored.

We can obtain some information about angular behavior of matrix
element from properties of the integral (\ref{matrelem}).
Obviously, we have periodicity of ${\cal M}$ in $\Phi$ with the
period $2\pi.$ For $\Phi=0$ and $\Phi=\pi$ we have immediately
${\cal M}=0.$ It follows from the oddness of the integrand for
these values of $\Phi.$ This investigation is a good check of
numerical results shown at Fig.~(\ref{fig:dsig_dPhi}). Of course,
more detailed information can be obtained only after numerical
integration of (\ref{matrelem}) with concrete functions
$f^{off}_{g,1},$ $f^{off}_{g,2}$ and $F_{g^*g^*\to\,\eta'}.$

The objects $f_{g,1}^{off}(x_1,x_1',k_{0,t}^2,k_{1,t}^2,t_1)$ and
$f_{g,2}^{off}(x_2,x_2',k_{0,t}^2,k_{2,t}^2,t_2)$ appearing in
formula (\ref{main_formula}) and (\ref{matrelem}) are skewed (or
off-diagonal) unintegrated gluon distributions. They are
non-diagonal both in $x$ and $k_t^2$ space. Usual off-diagonal
gluon distributions are non-diagonal only in $x$. In the limit
$x_{1,2} \to x_{1,2}'$, $ k_{0,t}^2 \to k_{1/2,t}^2$ and $t_{1,2}
\to 0$ they become usual UGDFs.

Using the relations (\ref{dec}) and $k_1-k_2=P_{M}$ and $s\gg
|{\bf k}_{0,t}|^2,$ we obtain
\begin{eqnarray}
s\,x_1x_2=m_{\eta'}^2+|{\bf p'}_{1,t}|^2+|{\bf p'}_{2,t}|^2+2|{\bf
p'}_{1,t}||{\bf p'}_{2,t}|\cos(\Phi)=m_{\eta'}^2+|{\bf
P}_{M,t}|^2\,.
\end{eqnarray}
The longitudinal momentum fractions are now calculated as:
\begin{eqnarray}
x_{1,2} &=& \frac{m_{\eta'}^2+|{\bf P}_{M,t}|^2}{s} \exp(\pm y)
 \nonumber \, , \\
x_{1,2}' &=& x_0= \frac{|{\bf k}_{0,t}|}{\sqrt{s}}   \, .
\label{x1_x2}
\end{eqnarray}
Above $y$ is the rapidity of the produced meson.

In the general case we do not know UGDFs very well. It seems
reasonable, at least in the first approximation, to take
\begin{eqnarray}
f_{g,1}^{off}(x_1,x_1',k_{0,t}^2,k_{1,t}^2,t_1) &=&
\sqrt{f_{g}^{(1)}(x_1',k_{0,t}^2) \cdot
f_{g}^{(1)}(x_1,k_{1,t}^2)} \cdot F_1(t_1)
\, , \\
f_{g,2}^{off}(x_2,x_2',k_{0,t}^2,k_{2,t}^2,t_2) &=&
\sqrt{f_{g}^{(2)}(x_2',k_{0,t}^2) \cdot
f_{g}^{(2)}(x_2,k_{2,t}^2)} \cdot F_1(t_2) \, ,
\label{skewed_UGDFs}
\end{eqnarray}
where $F_1(t_1)$ and $F_1(t_2)$ are usual Dirac isoscalar nucleon
form factors and $t_1$ and $t_2$ are total four-momentum transfers
in the first and second proton line, respectively. The above
prescription is a bit arbitrary,
although it is inspired by the positivity constraints \cite{posit}
for {\it collinear} Generalized Parton Distributions. 
It provides, however, an interpolation between different $x$ and
$k_t$ values appearing
kinematically. Our prescription is more symmetric in variables of
the first and second exchange than the one used recently in
\cite{BBKM06} for Higgs boson production.


The UGDFs above have a property that
\begin{eqnarray}
f(x,k_t^2) \to 0\, ,
\end{eqnarray}
if $k_t^2 \to$ 0. The small-$k_t^2$ region is of nonperturbative
nature and is rather modelled than derived from pQCD. Usually the
UGDFs in the literature are modelled to fulfill
\begin{eqnarray}
\frac{f(x,k_t^2)}{k_t^2} = {\cal F}(x,k_t^2) \to const
\label{kt2_limit}
\end{eqnarray}
if $k_t^2 \to$ 0. It is sometimes more useful to use ${\cal
F}(x,k_t^2)$ instead of $f(x,k_t^2)$.

When inspecting Eq.(\ref{main_formula}) and (\ref{skewed_UGDFs})
it becomes clear that the cross section for elastic
double-diffractive production of a meson (or Higgs boson) is much
more sensitive to the choice of UGDFs than the inclusive cross
sections.

\subsection{$\gamma^* \gamma^*$ fusion}

It was advocated in Ref.~\cite{CEF98} that the pseudoscalar mesons
production at small transverse momenta may be dominated by the
virtual photon -- virtual photon fusion. In the following we wish
to investigate the competition of the diffractive mechanism
discussed in the previous subsection and the $\gamma^* \gamma^*$
fusion mechanism.


\begin{figure}[h]     
 \centerline{\includegraphics[width=0.6\textwidth]{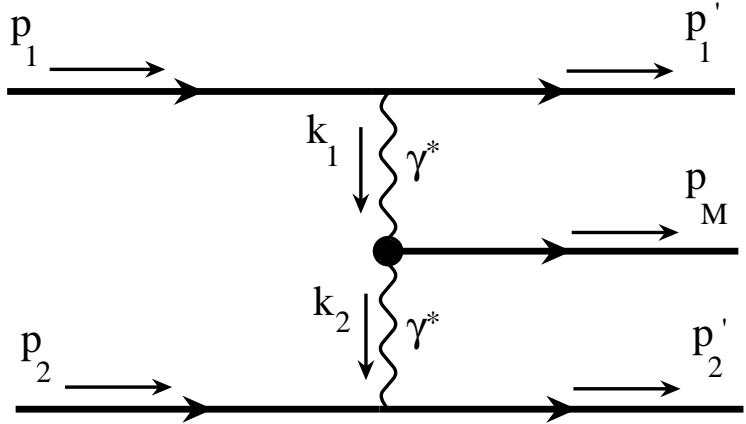}}
   \caption{\label{kinematics_gamgam}
   \small Kinematics of exclusive $\gamma^*\gamma^*$ fusion mechanism of $\eta'$-meson
 production.}
\end{figure}


In this case averaged matrix element squared
\begin{equation}
\overline { | {\cal M} |^2 } = \frac{1}{4} 
\sum |{\cal M}_{\lambda_1,\lambda_2,\lambda_1',\lambda_2'}|^2 \; .
\label{ME2_with_helicities}
\end{equation}
In the most general case the Born amplitude reads:
\begin{eqnarray}
{\cal M}_{\lambda_1,\lambda_2,\lambda_1',\lambda_2'}
= \{ \overline{u}(p'_1,\lambda_1')
[F_1(t_1) \gamma^{\mu}  \pm i\frac{\sigma^{\mu \mu''}}{2 M_N}
k_{1,\mu''} F_2(t_1) ] u(p_1,\lambda_1) \} \nonumber \\
 \frac{g_{\mu \mu'}}{t_1} \; \;
(- i) \, e^2 F_{\gamma^* \gamma^* \to \eta'}(t_1,t_2) 
\epsilon_{\mu' \nu' \alpha, \beta} 
k_1^{\alpha} k_2^{\beta} \; \;
 \frac{g_{\nu \nu'}}{t_2} \nonumber \\
\{ \overline{u}(p'_2,\lambda_2')
[F_1(t_2) \gamma^{\nu}  \pm i\frac{\sigma^{\nu \nu''}}{2 M_N}
k_{1,\nu''} F_2(t_2) ] u(p_2,\lambda_2) \}\; .  \nonumber \\
\label{QED_full_amplitude}
\end{eqnarray}
Limiting to large energies ($\sqrt{s} \gg m_{\eta'} + M_N + M_N$)
and small transverse momenta $t_1$ and $t_2$ ($|t| \ll 4 M_N^2$)
the matrix element for $pp \to p\eta'p$ reaction via virtual photon --
virtual photon fusion can be written as
\begin{eqnarray}
{\cal M}_{pp\, \to p\eta'p}^{\gamma^* \gamma^* \to\, \eta'}
\approx e F_1(t_1) \frac{(p_1 + p_1')^{\mu}}{t_1}
  \Gamma_{\mu\nu}^{\gamma^* \gamma^* \to\, \eta'}(k_1,k_2)
           \frac{(p_2 + p_2')^{\nu}}{t_2} e F_1(t_2)    \, ,
\label{gamma_gamma_eta_amplitude}
\end{eqnarray}
where
\begin{eqnarray}
\Gamma_{\mu\nu}^{\gamma^* \gamma^* \to\, \eta'}(k_1,k_2) = - i e^2
F_{\gamma^* \gamma^*\to\,\eta'}(k_1^2,k_2^2)\,\epsilon_{\mu \nu
\rho \sigma} k_1^{\rho} k_2^{\sigma}  \, .
\end{eqnarray}
In Eq.~(\ref{gamma_gamma_eta_amplitude}) $F_1(t_1)$ and $F_1(t_2)$
are Dirac proton electromagnetic form factors. In the following we have
omitted the spin-flipping contributions related to the respective Pauli
form factors. $F_{\gamma^* \gamma^* \to\, \eta'}$ is a respective
electromagnetic off-shell form factor normalized to
\begin{eqnarray}
F_{\gamma^* \gamma^* \to\, \eta'}(0,0) = \frac{1}{4 \pi^2
f_{\eta'}} \, ,
\end{eqnarray}
where $f_{\eta'}$ is the meson decay constant.
Alternatively one can use the relation
\begin{eqnarray}
|F_{\gamma^* \gamma^* \to\, \eta'}(0,0)|^2 = \frac{1}{(4 \pi
\alpha)^2} \frac{64 \pi \Gamma(\eta' \to \gamma \gamma)}
{m_{\eta'}^3} \, ,
\end{eqnarray}
where only measured quantities enter. Inserting PDG values of
experimental entries for $\eta'$ we get $|F_{\gamma^* \gamma^*
\to\, \eta'}(0,0)|^2$ = 0.116 GeV$^{-2}$.

Now we can write
\begin{eqnarray}
(p_1 + p_1')^{\mu} \Gamma_{\mu\nu}^{\gamma^* \gamma^* \to\,
\eta'}(k_1,k_2) (p_2 + p_2')^{\nu} \approx -ie^2 \cdot (2 s)
F_{\gamma^* \gamma^*\to\, \eta'}(k_1^2,k_2^2)\,|{\bf k}_{1,t}|
|{\bf k}_{2,t}| \sin(\Phi) \, . \label{gamma_gamma_eta_vertex}
\end{eqnarray}
Collecting all ingredients together we get
\begin{eqnarray}
\overline{|{\cal M}_{pp \to p\eta'p}^{\gamma^* \gamma^* \to\,
\eta'}|^2} \approx 4 s^2 e^8  \frac{F_1^2(t_1)}{t_1^2}
\frac{F_1^2(t_2)}{t_2^2} |F_{\gamma^* \gamma^*\to\,
\eta'}(k_1^2,k_2^2)|^2\, |{\bf k}_{1,t}|^2 |{\bf k}_{2,t}|^2
\sin^2(\Phi) \, . \label{gamma_gamma_amplitude_squared}
\end{eqnarray}

The $t_1$ and $t_2$ dependences of $F_{\gamma^* \gamma^* \to\,
\eta'}$ are the least known ingredients in the formula
 (\ref{gamma_gamma_amplitude_squared}).
It is known experimentally only for one virtual photon.
In the following we shall use two different forms of the form factors.
The first one is inspired by the vector-dominance model:
\begin{eqnarray}
F_{\gamma^* \gamma^* \to\, \eta'}(k_1^2,k_2^2) = \frac{F_{\gamma^*
\gamma^* \to\, \eta'}(0,0)}
{(1-k_1^2/m_{\rho}^2)(1-k_2^2/m_{\rho}^2)} \, .
\label{em_off_shell_formfactor1}
\end{eqnarray}
The second one is motivated by the leading twist pQCD analysis in
Ref.~\cite{DKV01}:
\begin{eqnarray}
F_{\gamma^* \gamma^* \to\, \eta'}(k_1^2,k_2^2) = \frac{F_{\gamma^*
\gamma^* \to\, \eta'}(0,0)}{(1-(k_1^2+k_2^2)/m_{\rho}^2)} \, .
\label{em_off_shell_formfactor2}
\end{eqnarray}
Both these forms describe the CLEO data \cite{CLEO} for one real
and one virtual photon.

\subsection{Phase space and kinematics}

The cross section for the 3-body reaction $p p \to p \eta' p$
can be written as
\begin{eqnarray}
d \sigma = \frac{1}{2 s} \overline { | {\cal M} |^2 } \cdot
d^{\,3} PS\, . \label{cross sect}
\end{eqnarray}
The three-body phase space volume element reads
\begin{eqnarray}
d^3 PS = \frac{d^3 p_1'}{2 E_1' (2 \pi)^3} \frac{d^3 p_2'}{2 E_2'
(2 \pi)^3} \frac{d^3 P_M}{2 E_M (2 \pi)^3} \cdot (2 \pi)^4
\delta^4 (p_1 + p_2 - p_1' - p_2' - P_M) \; . \label{dPS_element}
\end{eqnarray}
At high energies and small momentum transfers the phase space volume
element can be written as
\begin{eqnarray}
d^3 PS = \frac{1}{2^8 \pi^4} dt_1 dt_2 d\xi_1 d\xi_2 d \Phi \; \delta
\left( s(1-\xi_1)(1-\xi_2)-m_{\eta'}^2 \right) \; ,
\label{dPS_element_he1}
\end{eqnarray}
where $\xi_1$, $\xi_2$ are longitudinal momentum fractions carried by
outgoing protons with respect to their parent protons and
the relative angle between outgoing protons $\Phi \in (0, 2\pi)$.
Changing variables  $(\xi_1, \xi_2) \to (x_F, m_{\eta'}^2)$ one gets
\begin{eqnarray}
d^3 PS = \frac{1}{2^8 \pi^4} dt_1 dt_2 \frac{dx_F}{s \sqrt{x_F^2 +
4 (m_{\eta'}^2+|{\bf P}_{M,t}|^2)/s}} \; d \Phi \; .
\label{dPS_element_he2}
\end{eqnarray}

It is more convenient for lower (but still high) energy to use
variable $x_F$. However, at very high energies the cross section
becomes too much peaked at $x_F \approx$ 0 due to the jacobian
\begin{equation}
J \approx \frac{1}{\sqrt{x_F^2 + 4 m_{\eta'}^2/s}} \to \frac{\sqrt{s}}{2
  m_{\eta'}}
\label{jacobian}
\end{equation}
and the use of rapidity $y$ instead of $x_F$ is recommended.
The phase space element in this case has the following simple form
\begin{eqnarray}
d^3 PS = \frac{1}{2^8 \pi^4\,s} dt_1 dt_2 dy d \Phi\,.
\label{dPS_element_he3}
\end{eqnarray}
If $x_F$ is used then
\begin{equation}
\xi_{1,2} \approx 1 - \frac{1}{2} \sqrt{x_F^2 + \frac{4 m_{\eta'}^2}{s}}
\mp \frac{x_F}{2} \; .
\end{equation}
In the other case when the meson rapidity is used then
\begin{equation}
\xi_{1,2} \approx 1 - \frac{m_{\eta'}}{\sqrt{s}} \exp(\pm y) \; .
\end{equation}

Now the four-momentum transfers in both proton lines can be calculated as
\begin{eqnarray}
t_{1,2} = -\frac{p'^2_{1/2,t}}{\xi_{1,2}} - \frac{(1-\xi_{1,2})^2
m_p^2}{\xi_{1,2}} \; .
\label{t12_transverse_momenta}
\end{eqnarray}
Only if $\xi_{1,2}$ = 1, $t_{1,2} = -p'^2_{1/2,t}$. The latter
approximate relation was often used in earlier works on
diffractive production of particles. However, in practice $\xi_{1,2}
\ne$ 0 and the more exact equation must be used.
The range of $t_1$ and $t_2$ is not unlimited as it is
often assumed.
One can read off from Eq.(\ref{t12_transverse_momenta})
a kinematical upper limit for $t_{1,2}$ which is:
\begin{equation}
t_{1,2} < - \frac{(1-\xi_{1,2})^2}{\xi_{1,2}} m_p^2  \; .
\label{upper_limit_for_t12}
\end{equation}
In practice these phase space limits become active only
for $|x_F| >$ 0.2.
The lower limits are energy dependent but are not active in practice.

The Mandelstam variables for subsystems 
$\{\mathrm{proton}_1' + \eta'\}$ and
$\{\mathrm{proton}_2' + \eta'\}$ can be expressed via other
kinematical variables
\begin{eqnarray}
s_{1,2} = s(1-\xi_{1,2}) + m_p^2 + 2\, t_{1,2}  \; .
\end{eqnarray}
It is also checked if $s_{1,2} > (m_{\eta'} + m_p)^2$, but this limit is
not active in the region of interest (central production).

\section{Results}

Before we present our results let us discuss some input parameters
for our calculations.

The partial decay width $\Gamma(\eta' \to gg)$ is not well known.
Of course
\begin{equation}
\Gamma(\eta' \to gg) < \Gamma_{\eta'}^{tot} \approx 0.2\,
\mathrm{MeV} \; .
\end{equation}
In the following we shall take the upper limit in order to estimate
the cross section.

The form factors responsible for off-diagonal effects are taken
in the form
\begin{equation}
F(t_{1,2}) = \frac{4 m_p^2 - 2.79 t_{1,2}}
{(4 m_p^2 -t_{1,2})(1-t_{1,2}/071)^2} \; .
\label{off_diagonal_formfactors}
\end{equation}

The $k_1^2$ and $k_2^2$ dependence of the form factor
$F_{g^*g^*\to\,\eta'}(k_1^2,k_2^2)$ is not well known as it is due
to nonperturbative effects related to the internal structure of
the $\eta'$-meson. In the following, in analogy to the $\gamma^*
\gamma^* \to\, \eta'$ form factor, we take it in the factorized
double monopole form
\begin{equation}
F_{g^* g^* \to\, \eta'}(k_1^2,k_2^2) =
\frac{1}{(1-k_1^2/\Lambda_{os}^2)(1-k_2^2/\Lambda_{os}^2)} \;.
\label{off_shell_formfactor}
\end{equation}
We take $k_1^2 = - k_{1,t}^2$ and $k_2^2 = -k_{2,t}^2$. The
parameter $\Lambda_{os} \sim m_{\rho}$ may be expected. In general, it
can be treated as a free parameter in order to quantify the theoretical
uncertainties.

In the present work we shall use a few sets of unintegrated
gluon distributions which aim at the description of phenomena
where small gluon transverse momenta are involved.
Some details concerning the distributions can be found in Ref.\cite{LS06}.
We shall follow the notation there. We wish to stress in this context
that all the distributions considered give quite reasonable description of
the HERA $F_2$ data. Only scale dependent distributions require
a separate short discussion.

The Gaussian UGDF is obtained as
\begin{equation}
{\cal F}_{naive}(x,\kappa^2,\mu_F^2) = x g^{coll}(x,\mu_F^2)
\cdot f_{Gauss}(\kappa^2) / \pi \; ,
\label{naive_UGDF}
\end{equation}
Then the off-diagonal distribution is calculated as
\begin{eqnarray}
f_g^{off,1} = \sqrt
{
f_g^{1}(x_1',k_{0t}^2,\mu_1^2) f_g^{2}(x_1,k_{1t}^2,\mu_2^2)
}
\cdot F_1(t_1) \nonumber \\
f_g^{off,2} = \sqrt
{
f_g^{1}(x_2',k_{0t}^2,\mu_1^2) f_g^{2}(x_2,k_{1t}^2,\mu_2^2)
}
\cdot F_1(t_1)
\label{off_diagonal_gauss}
\end{eqnarray}
The choice of the (factorization) scale here is not completely obvious.
We shall try two choices:\\
(a) $\mu_1^2 = m_{\eta'}^2, \mu_2^2 = m_{\eta'}^2 \; ,$ \\
(b) $\mu_1^2 = Q_0^2, \mu_2^2 = m_{\eta'}^2 \;.$ \\
The first choice is similar as in \cite{KMR}.
However, it is not obvious if the scale associated with the ``hard''
production ($g^* g^* \to \eta'$) can be used for the left part
of the gluonic ladder where no obvious hard scale appears.
Therefore we shall try also the second choice where we shall use
$Q_0^2$ = 0.26 GeV$^2$, i.e. the nonperturbative input for the QCD
evolution in Ref. \cite{GRV95}.

\begin{figure}[!h]    
 \centerline{\includegraphics[width=0.5\textwidth]{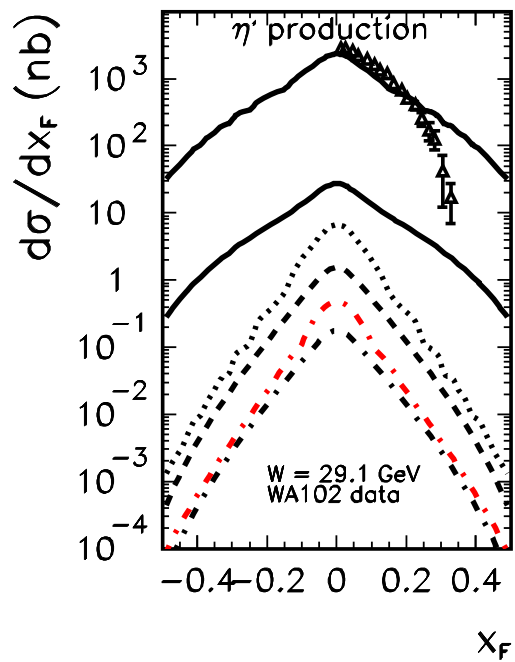}}
   \caption{\label{fig:dsig_dxF}
   \small $d \sigma / dx_F$ as a function of Feynman $x_F$ for W
= 29.1 GeV and for different UGDFs.
The $\gamma^* \gamma^*$ fusion contribution is shown by
the dash-dotted (red) line (second from the bottom).
The experimental data of the WA102 collaboration \cite{WA102} are shown
for comparison.}
\end{figure}

Let us start from the $d\sigma/dx_F$ distribution. In
Fig.~\ref{fig:dsig_dxF} we show the results of calculations
obtained with several models of UGDF (for details see \cite{LS06}).
For comparison we show also the contribution of the
$\gamma^* \gamma^*$ fusion mechanism.
The contribution of the last mechanism is much smaller than
the contribution of the diffractive QCD mechanism.

\begin{figure}[!h]     
 \centerline{\includegraphics[width=0.45\textwidth]{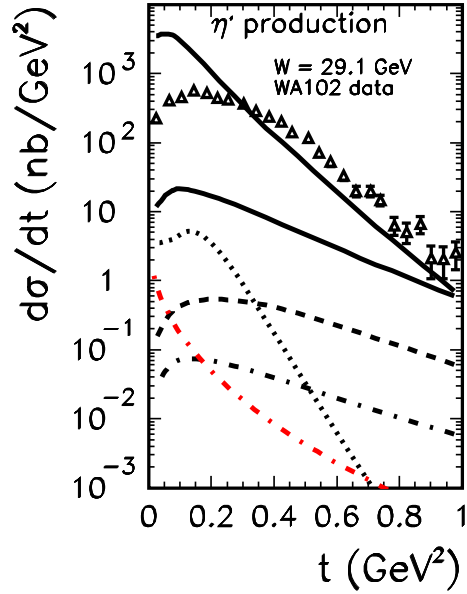}}
   \caption{ \label{fig:dsig_dt12}
\small $d \sigma / dt_{1/2}$ as a function of Feynman $t_{1/2}$
for W = 29.1 GeV and for different UGDFs.
The $\gamma^* \gamma^*$ fusion contribution is shown by the dash-dotted
(red) steeply falling down line.
The experimental data of the WA102 collaboration \cite{WA102} are shown
for comparison.}
\end{figure}

In Fig.~\ref{fig:dsig_dt12} we present distribution in $t_1$ and
$t_2$ (identical) of the diffractive production and
of the $\gamma^* \gamma^*$ mechanism (red dash-dotted curve). The
distribution for the $\gamma^* \gamma^*$ fusion is much steeper
than that for the diffractive production.

\begin{figure}[!h]       
 \centerline{\includegraphics[width=0.45\textwidth]{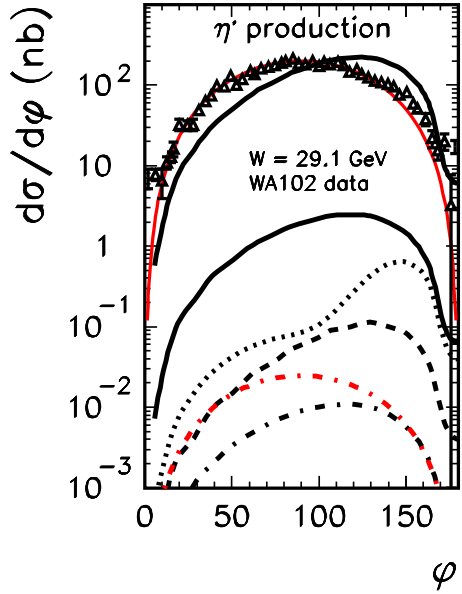}}
   \caption{ \label{fig:dsig_dPhi}
\small  $d \sigma / d\Phi$ as a function of $\Phi$ for W = 29.1
GeV and for different UGDFs.
The $\gamma^* \gamma^*$ fusion contribution is shown by the dash-dotted
(red) symmetric around 90$^o$ line.
The experimental data of the WA102 collaboration \cite{WA102} are shown
for comparison.}
\end{figure}

In Fig.~\ref{fig:dsig_dPhi} we show the distribution of the cross
section as a function of the angle between the outgoing protons.
In the first approximation it reminds $\sin^2(\Phi)$. A
more detailed inspection shows, however, that the distribution is
somewhat skewed with respect to $\sin^2(\Phi)$ dependence.
This is due to the two reasons:\\
(a) kinematical -- caused by interrelations of integration variables
    due to finite phase-space limits
    (present also for the pomeron + pomeron  $\to \eta'$ fusion model), \\
(b) dynamical -- caused by nonlocality due to the internal loop
    of the diagram shown in Fig.~\ref{fig:diffraction_updf}
    (the $sin(\phi)$ dependence is embedded only in the loop integration).\\
In order to quantify the effect we define the parameter of
the skewedness of the $\Phi$ distribution as
\begin{equation}
S_{\pi/2}(W) \equiv
\frac{   \int_{\pi/2}^{\pi} \frac{d \sigma}{d \Phi}\left( W \right) d \Phi
       - \int_{0}^{\pi/2} \frac{d \sigma}{d \Phi}\left( W \right) d \Phi     }
{ \int_{0}^{\pi} \frac{d \sigma}{d \Phi}\left( W \right) d \Phi }  \; .
\label{skewdness_parameter}
\end{equation}
If we take more differential cross section in the definition above than
$S_{\pi/2} = S_{\pi/2}(W,y,t_1,t_2)$ (or $S_{\pi/2}(W,x_F,t_1,t_2)$).
Of course -1 $< S_{\pi/2} <$ 1.
For exact $\sin^2(\Phi)$ distribution $S_{\pi/2}$ = 0. In Table 1
we show the skewedness parameter $S_{\pi/2}$ for our model for
different initial energies $W =$  29.1, 200, 1960, 14000 GeV, relevant
for WA102, RHIC, Tevatron and LHC, respectively.
Generally, the larger energies the smaller $S_{\pi/2}$.
The phase space limitations cause only a very small skewedness.



%
%
%
%



\begin{table}
\caption{
The measure of the skewedness $S_{\pi/2}$ of azimuthal angle
distributions for different UGDFs and different center-of-mass energies.
In this calculation -0.5 $< x_F <$ 0.5, -1 GeV $< t_{1,2} <$ 0.
}
\begin{center}

\begin{tabular}{|c|c|c|c|c|}
\hline W (GeV) & $S_{\pi/2}$ (KL) & $S_{\pi/2}$ (GBW) & $S_{\pi/2}$
(BFKL)  \\
\hline
 29.1  & 0.5990 & 0.7889 & 0.3615  \\
200    & 0.5867 & 0.6628 & 0.3131  \\
500    & 0.5629 & 0.4983 & 0.2990  \\
1960   & 0.5019 & 0.2622 & 0.2814  \\
14000  & 0.3870 & 0.2283 & 0.2617  \\
\hline
\end{tabular}

\end{center}

\end{table}


\begin{table}

\caption{Energy dependence of the cross section (in nb)
for different UGDFs.
The integration is over -4 $< y <$ 4 and -1 GeV $< t_{1,2} <$ 0.
The second lines for Gaussian distributions are for the choice (b)
of the factorization scale.
No absorption corrections were included.
}

\begin{center}

\begin{tabular}{|c|c|c|c|}
\hline
UGDF & 29.1 & 200 & 1960 \\ 
\hline
 KL                 & 0.2867(+0) & 0.7377(+0) & 0.4858(+0) \\
 GBW                & 0.1106(+1) & 0.2331(+2) & 0.1034(+3) \\
 BFKL               & 0.3279(-1) & 0.9205(+1) & 0.2188(+4) \\
 Gauss (0.2)        & 0.6391(+3) & 0.1697(+5) & 0.2964(+6) \\
                    & 0.3445(+1) &            & 0.2984(+3) \\
 Gauss (0.5)        & 0.7389(+1) & 0.2705(+3) & 0.3793(+3) \\ 
                    & 0.4199(-1) &            & 0.4094(+1) \\ 
\hline
$\gamma^* \gamma^*$ & 0.7764(-1) & 0.2260(+0) & 0.3095(+0) \\
\hline
\end{tabular}

\end{center}

\end{table}



\begin{figure}[!h]       
 \centerline{\includegraphics[width=0.60\textwidth]{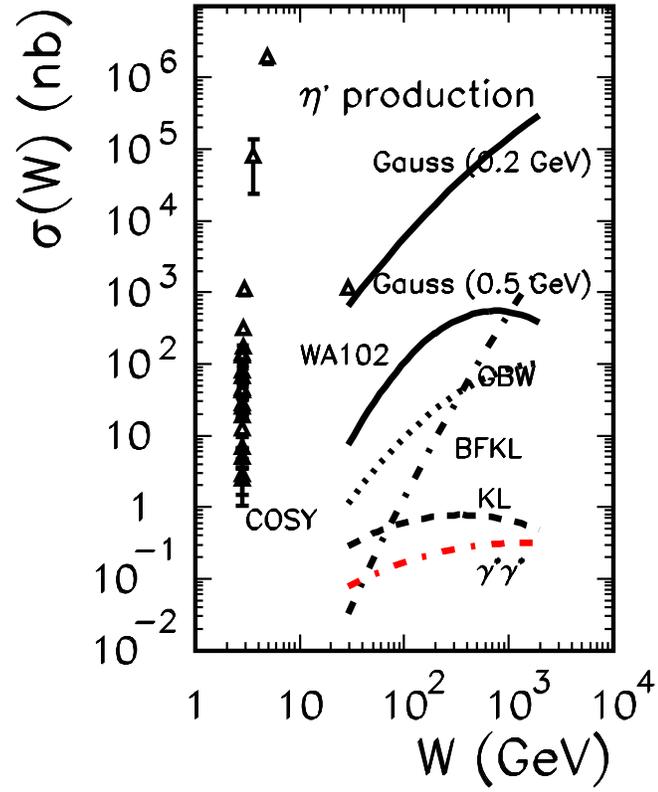}}
   \caption{ \label{fig:sig_tot_w}
\small  $\sigma_{tot}$ as a function of center of mass energy
for different UGDFs.
The $\gamma^* \gamma^*$ fusion contribution is shown by the dash-dotted
(red) line. The world experimental data are shown for reference.}
\end{figure}

In Table 2 and in Fig.\ref{fig:sig_tot_w} we show energy dependence of
the total (integrated over kinematical variables) cross section for
the exclusive reaction $p p \to p \eta' p$ for different UGDFs.
In the case of Gaussian UGDFs we show in Table 2 also results
with the second choice of the factorization scale.
The cross section with the second choice is much smaller than
the cross section with the first choice.
Quite different results are obtained for different UGDFs.
This demonstrates once again the huge sensitivity to the choice of UGDF.
The cross section with the Kharzeev-Levin type distribution (based
on the idea of gluon saturation) gives
the cross section which is small and almost idependent of beam energy.
In contrast, the BFKL distribution leads to strong energy dependence.
The sensitivity to the transverse momenta of initial gluons
can be seen by comparison of the two solid lines calculated with
the Gaussian UGDF with different smearing parameter
$\sigma_0$ = 0.2 and 0.5 GeV.
The contribution of the $\gamma^* \gamma^*$ fusion mechanism 
(red dash-dotted line) is fairly small and only slowly energy dependent.
While the QED contribution can be reliably calculated, the QCD
contribution cannot be at present fully controlled.
It is even not completely excluded that the QED contribution dominates
over the QCD contribution in some energy window.

\begin{figure}[!hp]       
\begin{minipage}{0.49\textwidth}
\epsfxsize=\textwidth\epsfbox{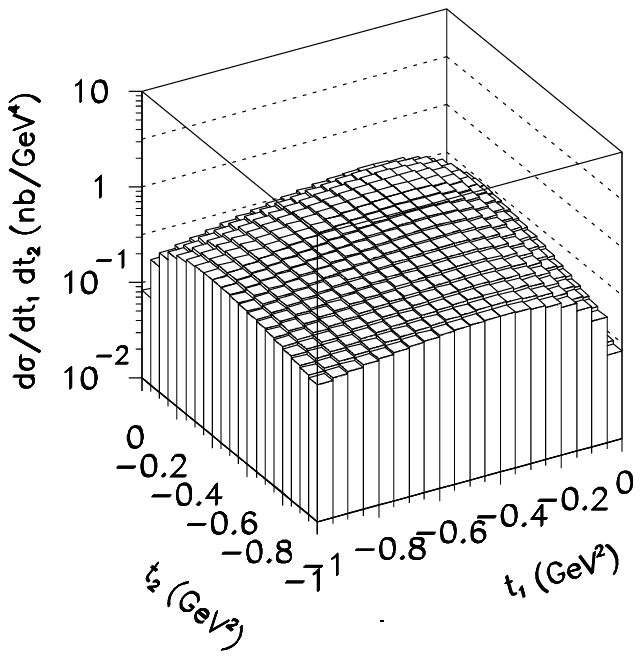}
\end{minipage}
\begin{minipage}{0.49\textwidth}
\epsfxsize=\textwidth \epsfbox{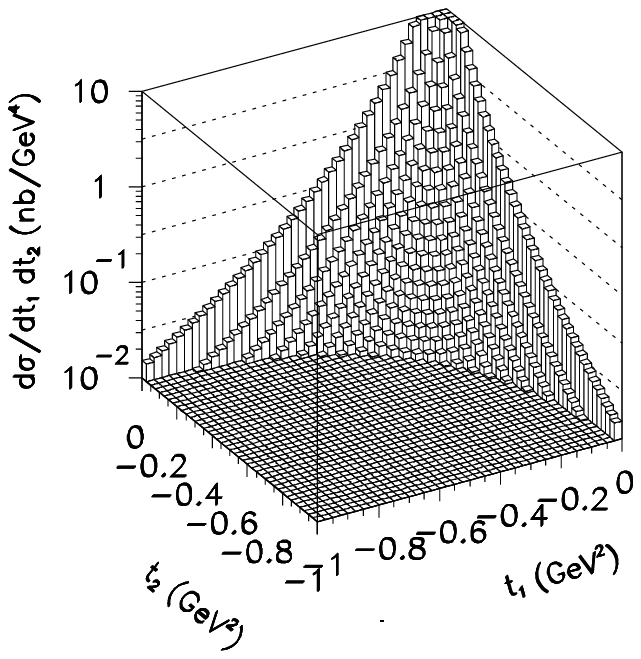}
\end{minipage}
\caption{\small Two-dimensional distribution in $t_1 \times t_2$
for the diffractive QCD mechanism (left panel), calculated with the KL
UGDF, and the $\gamma^* \gamma^*$ fusion (right panel) at
the Tevatron energy W = 1960 GeV.}
\label{fig:map_t1t2}
\end{figure}

At present it seems impossible to understand the dynamics of
the exclusive $\eta'$ production at high energy without a real
measurement.
The Tevatron apparatus gives such a possibility, at least in principle.
In Fig.\ref{fig:map_t1t2} we present two-dimensional maps
$t_1 \times t_2$ of the cross section for the QCD mechanism (KL UGDF)
and the QED mechanism (Dirac terms only) for the Tevatron energy W = 1960 GeV.
If $ |t_1|, |t_2| > $ 0.5 GeV$^2$ the QED mechanism is clearly negligible.
However, at $|t_1|, |t_2| < $ 0.2 GeV$^2$ the QED mechanism may become
equally important or even dominant. In addition, it may interfere with
the QCD mechanism.

In Fig.\ref{fig:map_tphi} we show a two-dimensional map $t \times \Phi$,
where $t = t_1$ or $t_2$. The bigger $t$, the larger skewedness with
respect to $\Phi = \pi/2$. The skewedness, which is almost independent
of the beam energy, is a generic feature of the QCD mechanism,
quite independent of the choice of UGDF. The observation of
the skewedness seems to be a condition ``sine qua non'' for
the confirmation of the QCD mechanism.
\footnote{The absorption corrections should only increase the
  skewedness.}
On the other hand, the observation of the $\sin^2 \Phi$ dependence,
as for the lower-energy WA102 data, may be very difficult to
understand microscopically.

Summarizing, the reaction under consideration seems very promissing
in better understanding of the QCD dynamics in the nonperturbative region.

In the case of Higgs (or heavy particle) production the large mass
sets a hard scale. In the case of $\eta'$ the mass is only about 1 GeV.
In this case the hard scale can be obtained by selecting large
transverse momenta or analogously large $|t_1|$ and $|t_2|$.
In principle, these variables could be controlled by measuring outgoing
protons.
As an example in Table 3 we have collected results for different
windows in the $(t_1, t_2)$ space for the KL UGDF.
In this case the cross section is dropping down rather slowly
with increasing $|t_1|$ and $|t_2|$.
However, this result depends strongly on UGDF used in the calculation.
Therefore measuring the cross section for different cuts
on $t_1$ and $t_2$ would be a farther test of UGDFs.


\begin{table}

\caption{Cross section in nb for different cuts on $t_1$ and $t_2$
and -0.5 $< x_F <$ 0.5.
The limits for the $(-t_1)$ and $(-t_2)$ windows in GeV$^2$ are given
explicitly. This calculation was done for the KL UGDF. 
}

\begin{center}

\begin{tabular}{|c|c|c|c|c|}
\hline
        &   (0,1)    &   (1,2)    &   (2,3)    &   (3,4)    \\
\hline
 (0,1)  & 0.1587( 0) & 0.5437(-1) & 0.8436(-2) & 0.1558(-2) \\
 (1,2)  & 0.5437(-1) & 0.2257(-1) & 0.4033(-2) & 0.7978(-3) \\
 (2,3)  & 0.8436(-2) & 0.4033(-2) & 0.9033(-3) & 0.2089(-3) \\
 (3,4)  & 0.1558(-2) & 0.7978(-3) & 0.2089(-3) & 0.5753(-4) \\
\hline

\end{tabular}

\end{center}

\end{table}



\begin{table}

\caption{Comparison of the cross section (in nb) for $\eta'$
and $\eta_c$ production at Tevatron (W = 1960 GeV)
for different UGDFs.
The integration is over -4 $< y <$ 4 and -1 GeV $< t_{1,2} <$ 0.
The second lines for Gaussian distributions are for the choice (b)
of the factorization scale.
No absorption corrections were included.
}

\begin{center}

\begin{tabular}{|c|c|c|c|}
\hline
UGDF & $\eta'$ & $\eta_c$ \\ 
\hline
 KL                 & 0.4858(+0)       & 0.7392(+0) \\
 GBW                & 0.1034(+3)       & 0.2039(+3) \\       
 BFKL               & 0.2188(+4)       & 0.1618(+4) \\
 Gauss (0.2)        & 0.2964(+6)       & 0.3519(+8) \\
                    & 0.2984(+3)       & 0.2104(+4) \\
 Gauss (0.5)        & 0.3793(+3)       & 0.4417(+6) \\
                    & 0.4094(+1)       & 0.3008(+2) \\
\hline
$\gamma^* \gamma^*$ & 0.3095(+0)       & 0.4493(+0) \\
\hline
\end{tabular}

\end{center}

\end{table}


The formalism presented in the previous sections can be applied to a
production of other pseudoscalar mesons.
In Table 4 we have collected cross sections for $\eta_c$ meson
integrated over broad range of kinematical variables specified in
the table caption. Again we have taken an upper limit assuming
$\Gamma(\eta_c \to gg) = \Gamma_{\eta_c}^{tot}$, which may be even more
reliable in the case of $\eta_c$ production.
These cross sections are very similar to the cross section for $\eta'$
production and in some cases even bigger. The results with Gaussian
distribution, $\sigma_0$ = 0.2 GeV and first choice of factorization
scale seems excluded,
as constituting too large fraction of the total cross section.
This strongly suggests also that the analogous result for $\eta'$
production must be questioned. This seems to open a problem of
understanding the WA102 data in terms of the QCD mechanism discussed
above.

Our result shows that the measurement of double-diffractive
double-elastic production of $\eta_c$ should be possible.
However, one should remember about very small branching fractions
for different decay channels of $\eta_c$. It is not clear to us
at present if the missing mass technique could be used at the Fermilab 
Tevatron. This would help to avoid the small branching fraction problem.
The results obtained with different UGDFs differ significantly.
Any measurement of the reaction would be then very interesting to
estimate (or limit) UGDFs in the nonperturbative region.

\section{Discussion and Conclusions}

For the first time in the literature, we have calculated exclusive
production of $\eta'$ meson in high-energy $pp \to p\eta'p$
collisions within the formalism of unintegrated gluon
distributions. This type of reaction exhibits an incredible
sensitivity to the choice of UGDF, which makes precise predictions
rather difficult. Measurements of this reaction, however, would
help to limit or even pin down the UGDFs in the nonperturbative
region of small gluon transverse momenta $k_t$ where these objects
cannot be obtained as a solution of any perturbative evolution equation,
but must be rather modelled. The usual procedure is to extrapolate the
perturbative regime via a smooth parametrization. For most of
inclusive reactions the details of such a procedure are not
essential. In contrast, for the reaction discussed here the
extrapolation is crucial.

The existing models of UGDFs predict cross section much smaller
than the one obtained by the WA102 collaboration at the center-of-mass
energy W = 29.1 GeV. This may signal presence of subleading reggeons
at the energy of the WA102 experiment or suggest a modificaction of UGDFs
in the nonperturbative region of very small transverse momenta.
Experiments on exclusive central production of $\eta$' at RHIC,
Tevatron and LHC would certainly help in disentangling the problem.
With some cuts on $\xi_1$, $\xi_2$, $t_1$ and $t_2$ the reaction
under consideration can be measured at the Fermilab Tevatron \cite{Royon}.
An exact evaluation of the experimental cross section for the Tevatron
will be presented elsewhere.

A reasonable description of the WA102 total cross section can be
obtained with UGDF obtained by Gaussian smearing of collinear gluon
distributions and rather small value of the smearing parameter
$\sigma_0 \sim$ 0.2 - 0.3 GeV, clearly pointing to a nonperturbative
effect. If the parameter $\sigma_0$ is adjusted to the total cross
section a reasonable description of the $d\sigma/dx_F$ around
$|x_F| <$ 0.2  is obtained simultaneously. This was not possible with
the Regge-like (two-pomeron exchange) description of the reaction \cite{KMV99}
which produced distribution too much peaked at $x_F \approx$ 0.
However, our approach gives somewhat too much skewed (asymmetric
around $\Phi = \pi/2$) distribution
in relative azimuthal angle between outgoing protons compared
to the WA102 data. This model gives definite predictions
at larger energies where the contribution of subleading reggeons
should be negligible.
Experimental data at different collision energies
would verify the solution and shed more light on the dynamics
of the $\eta'$ meson production. At present the Tevatron apparatus
could be used.

Measurement at lower energies would be also interesting.
Natural possibilities would be FAIR at GSI and J-PARC at Tokai.
Such data could shed more light on the role of subleading reggeons
which seems important in understanding the WA102 data. 

Due to a nonlocality of the loop integral our model leads to sizeable
deviations from the $\sin^2 \Phi$ dependence (predicted in the models
of one-step fusion of two vector objects).

The $\gamma^* \gamma^*$ fusion gives the cross section of the order
of a fraction of nb at the WA102 energy W = 29.1 GeV, i.e. much less
than 1 \% of the measured cross section. The $\gamma^* \gamma^*$ fusion
may be of some importance only at extremely small four-momentum transfers
squared. In addition it can interfere with the QCD mechanism, which is
similar to the familiar Coulomb-nuclear interference for charged hadron
elastic scattering. 

Finally we have presented results for exclusive double elastic
$\eta_c$ production. Similar cross sections as for $\eta'$ production
were obtained. Also in this case the results depend strongly on
the choice of UGDF.

In the present calculations we have calculated only so-called 
bare amplitudes which are subjected to absorption corrections.
The absorption effects lead usually to an energy-dependent damping
of the cross section for exclusive channels. At the energy of
the WA102 experiment W = 29.1 GeV the damping factor is expected to be of
the order 5 -- 10 and should increase with rising initial energy.
A detailed analysis of absorption effects is left for a future separate
study.

\vskip 0.5cm

{\bf Acknowledgements} We are indebted to Kolya Kochelev and Andrey
Vinnikov for providing us the WA102 data and Pawe{\l} Moskal
for providing us low-energy data.
We also thank Christophe Royon for a discussion about possibilities
to measure the $p p \to p \eta' p$ reaction at the Fermilab Tevatron.
We are grateful to Alexei Kaidalov, Valery Khoze and Roland Kirschner
for a discussion on the applicability of QCD factorization to
double diffraction.  
This work was partially supported by the grant
of the Polish Ministry of Scientific Research and Information Technology
number 1 P03B 028 28 and by RFBR grant 06-02-16215.


\newpage


\begin{figure}[!h]       
 \centerline{\includegraphics[width=0.60\textwidth]{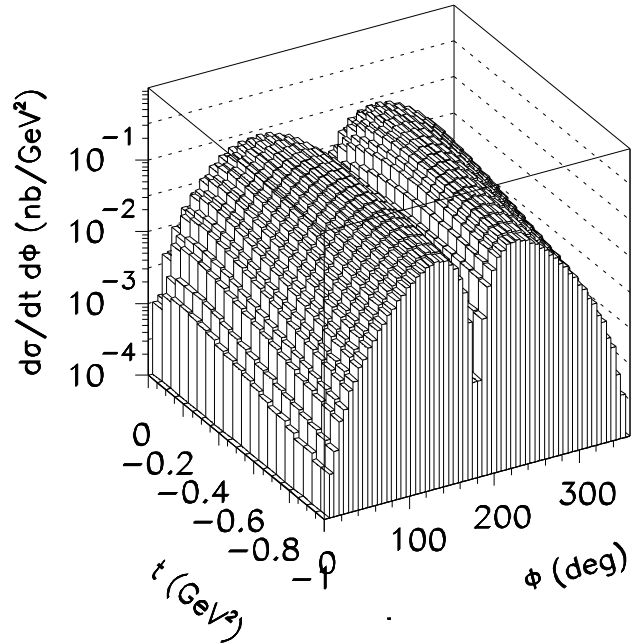}}
   \caption{ \label{fig:map_tphi}
\small  The cross section for $ p \bar p \to p \eta' \bar p $
as a function of $t$ ($t_1$ or $t_2$)
and relative azimuthal angle $\Phi$ for the Tevatron energy
W = 1960 GeV. In this calculation the KL UGDF was used.
}
\label{fig:map_tphi}
\end{figure}


\end{document}